\documentclass[%
reprint,showkeys,
nofootinbib,  
amsmath,amssymb,
aps]{revtex4-2}

\usepackage{graphicx}
\usepackage{dcolumn}
\usepackage{bm}
\usepackage{siunitx}
\usepackage{xcolor}
\usepackage{balance}
\usepackage[english]{babel}
\usepackage{bbding}
\usepackage[version=3]{mhchem}
\usepackage{mathptmx,amsmath,amssymb}

\begin{document}
	\title{Friction through molecular adsorption at the sliding interface of hydrogels:\\ Theory and experiments\\}



	\begin{abstract}
		{Lola Ciapa, Ludovic Olanier, Yvette Tran, Christian Fr\'etigny, Antoine Chateauminois, Emilie Verneuil \\
			{\it \noindent Soft Matter Science and Engineering (SIMM), CNRS UMR 7615, ESPCI Paris, PSL University, Sorbonne Universit\'e, F-75005 Paris, France\\
			antoine.chateauminois@espci.fr, emilie.verneuil@espci.fr\\}}

	We report on the frictional properties of thin ($\approx \mu m$) poly(dimethylacrylamide) hydrogel films within contacts with spherical silica probes. In order
to focus on the contribution to friction of interfacial dissipation, a dedicated rotational setup is designed which allows to suppress poroelastic flows while ensuring an uniform velocity field at the sliding interface. The physical-chemistry of the interface is varied from the grafting of various silanes on the silica probes. Remarkably, we identify a velocity range in which the average frictional stress systematically varies with the logarithm of the sliding velocity. This dependency is found to be sensitive to the physical-chemistry of the silica surfaces. Experimental observations are discussed in the light of a molecular model where friction arises from thermally activated adsorption of polymer chains at the sliding interface, their elastic stretching and subsequent desorption. From this theoretical description, our experimental data provide us with adhesion energies and characteristic times for molecular adsorption that are found consistent with the physico-chemistry of the chemically-modified silica surfaces.

\end{abstract}

	\maketitle
\footnotetext{\dag~Electronic Supplementary Information (ESI) available: SI1: Scaling of friction force with contact radius; SI2: Frictional shear stress in the low velocity regime.]. 
}


\section{Introduction}

Hydrogels consist of hydrophilic polymer networks that can absorb large quantities of water. As such, they have recently been developed as promising materials for varied applications such as optical engineering, bioengineering, coatings, microfluidics,... When used as substitute for cartilages,\cite{porte_lubrication_2020,blum_low_2013,han_effect_2018} as contact lenses \cite{dunn_lubricity_2013,roba_friction_2011} or as anti-fog coatings for windows and windshields,\cite{delavoipiere_swelling_2018, lee_zwitter-wettability_2013} a good control of their frictional properties is needed. Recently, attempts have been made to tune the frictional properties of hydrogels by changing the physico-chemistry of their surface or of the counter-surface. Hence, among other examples, changing the counter surface hydrophobicity \cite{tominaga_effect_2008} or the electrostatic charges at the two interfaces \cite{kagata_surface_2001} was found to change both the shape and the magnitude of the friction-velocity curves. However, the friction force was found to result from at least two intricate mechanisms: interfacial friction and viscous dissipation. Interfacial friction is reminiscent of the Schallamach model for rubber friction \cite{schallamach_theory_1963}: it results from the molecular adsorption, stretching and desorption of the gel polymers at the interface. Viscous dissipation occurs within a thin lubrication film of solvent possibly squeezed at the sliding interface.\cite{skotheim_soft_2005, salez_elastohydrodynamics_2015} While the former is expected to dominate at low velocity, the latter arises at larger velocity where lubrication is forced, \cite{gong_gel_1998} but their relative weight still lacks a quantitative description. This is due to two especially challenging experimental characterizations:
the molecular parameters controlling the adsorption mechanisms across the confined interface, \cite{xiang_surface_2020} and the thickness\cite{gong_gel_1998, yamamoto_situ_2014, simic_importance_2020} and rheological behavior\cite{baumberger_self-healing_2003} of a lubrication film when it exists.\cite{kagata_friction_2002, kurokawa_elastichydrodynamic_2005,cuccia_pore-size_2020} In addition, a third effect comes into play when contacts are displaced at the hydrogel surface: poroelasticity. Indeed, the normal load translates into a porous flow within the gel network until the load balances with the elastic response of the polymer network. These flows result in a velocity- and time-dependent contact area \cite{delavoipiere_friction_2018, ciapa_transient_2020} and possibly extra frictional dissipation. Hence, in friction experiments where a macroscopic solid or an AFM probe slides on a hydrogel \cite{shoaib_insight_2018}, the three mechanisms should {\it a priori} operate simultaneously. Hence, understanding hydrogel friction still lacks a precise description of each one of the dissipative phenomena, and of their possible couplings.\\

In the present paper, we offer to gain insights into the physical mechanisms at play in hydrogel friction by designing experiments aiming at isolating the molecular adsorption/desorption mechanisms and measuring their molecular parameters (adsorption energy and time, surface density). To do so, we will show that our experimental set-up and system allows to minimize lubrication and poroelastic transport so that the friction-velocity data we measure can be interpreted solely in terms of interfacial friction as defined above. In the past, a large variety of experiments have been developed in order to measure polymer/substrate interactions, at both macroscopic and single molecule scales. Macroscopic tests include static and dynamic adhesion experiments\cite{johnson_surface_1971}, flows of polymer melts\cite{henot_temperature_2018}, or friction on elastomers\cite{henot_friction_2018} or glassy polymers\cite{bureau_friction_2007}. Their interpretation is not always straightforward due to complex couplings between interfacial and bulk dissipations, non-uniform interfacial velocities, or possibly collective effects. 
 {In an alternate and interesting way to investigate the role of pinning/depinning molecular processes on friction, Bennewitz and collaborators designed functionalised silica surfaces whose frictional and adhesive properties were tuned through switchable guest/host interactions by applying either an electric potential \cite{bozna_friction_2015} or light \cite{blass_switching_2015}.}
 {Single-molecule experiments have been helpful at assessing the molecular parameters governing the break of single bonds in pull-out experiments\cite{friddle_near-equilibrium_2008} or friction of single molecules using AFM tips\cite{kuhner_friction_2006}. The force-extension spectra measured in these works - sometimes combining both in- and off-plane displacements- were successfully interpreted with models based on thermally activated force-induced desorption to obtain molecular adhesion energies or molecular mobility. The dependency of these adsorption energies with extension rate\cite{friddle_near-equilibrium_2008} or orientation of the imposed displacement\cite{kuhner_friction_2006} remains however unclear.}
 {
Finally, friction between two monolayers of C12-alkyl chains obtained by the strong adsorption of cationic surfactants on mica was measured using a Surface Force Apparatus nanotribometer over decades in sliding velocities \cite{drummond_friction_2003} and convincingly interpreted with a thermally activated model derived from Schallamach.\cite{schallamach_theory_1963} The authors however point out that (i) for such dense monolayers, correlation between neighboring bonds cannot be avoided and (ii) for these short alkyl chains for which the in-plane and out-of-plane characteristic lengths are close, the bond stiffness and the density of bonds cannot be separated so that the dependence with the surfactant concentration could not be elucidated.  }

 {Here, by working on friction of hydrogels, we expect that the relatively low density of active polymer chains at the interface will support the hypothesis of independent bonds, while the stiffness of the bond will be set by the length of the interfacial chains: this enhanced compliance will allow for relatively larger displacements before break that can be modelled. We will show that the molecular adhesion, the typical adsorption times, and the surface density involved in an irreversible adsorption-stretching-desorption dissipative mechanism can be measured between a variety of solid silica surfaces functionalised with different molecules and a swollen hydrogel.}
%
\section{Material and methods}
\subsection{Hydrogel films}\label{sec:hydrogel}
All the experiments to be reported were carried out using poly(dimethylacrylamide) (PDMA) hydrogel films covalently grafted onto glass substrates. These films were synthesized by simultaneously crosslinking and grafting preformed ene-functionalised polymer chains onto glass substrates using a thiol-ene click reaction which is fully described elsewhere \cite{delavoipiere_poroelastic_2016, delavoipiere_friction_2018} and allows to separately control the film thickness and swelling ratio. Strong adhesion between the hydrogel film and its substrate was achieved by functionalising borosilicate glass slides with thiol groups. Through a thiol-ene reaction, the gel film was covalently bound to the glass slide, thereby preventing interfacial debonding during swelling and friction. In order to assess the reproducibility of our results, two different PDMA films with slightly different thicknesses and swelling ratio were synthesized and their dry and swollen thicknesses were measured by spectroscopic ellipsometry. The dry thickness $e_0$ and swelling ratio $Sw$ are $e_0=2.8$~\si{\micro\meter} and $Sw=2.3$ for one film, and $e_0=2.3$~\si{\micro\meter} and $Sw=1.8$ for the other one.  {The elastic response of these two films was measured by indentation experiments using a procedure published elsewhere \cite{delavoipiere_poroelastic_2016}. For normal loads up to 200~mN, the indentation curve was successfully described using a linear stress/strain relationship characterized by an elastic oedometric modulus $\tilde{E}=21$~MPa and $30$~MPa respectively. Finally, the shear modulus of the drained network was deducted\cite{augustine_swelling_2023} from the swelling ratio $Sw$ to be $G_0=1.9$~MPa and $4.0$~MPa respectively. From the shear modulus, we estimate the number of Kuhn segments between crosslinks $\nu_c=6$ and $\nu_c=13$ respectively, so that $\nu_c$ will be approximated as $\nu_c\sim 10$ in what follows. The large value of these moduli, together with the high crosslink density, were chosen to avoid film damage upon swelling \cite{augustine_swelling_2023} and sliding \cite{delavoipiere_friction_2018, ciapa_transient_2020}.}
\subsection{Functionalised silica surfaces}\label{sec:silica:charac}
As solid probe, we used fused silica lenses (Newport, UV fused silica SPX114) with curvature radius $R=23$~mm. In order to vary their molecular interactions with PDMA hydrogel, the silica surface was grafted with three different silanes, namely aminopropyltriethoxysilane (APTES, ABCR GmbH), propyltriethoxysilane (PTES, ABCR GmbH) and octadecyltrichlorosilane (OTS, ABCR GmbH).\\
The silica surface is first activated with a freshly prepared piranha (H$_2$SO$_4$/H$_2$O$_2$) solution at 150~\si{\celsius}, rinsed and sonicated in Milli-Q water
prior to drying under nitrogen flow and subsequent UV/ozone treatment. Then, the freshly cleaned silica lenses are quickly transferred into a sealed reactor filled with nitrogen where a solution of dry toluene with the silane is introduced. For APTES and PTES silanes, the lenses are soaked for three hours in solutions containing 5\%vol and 3\%vol of silane, respectively. For OTS, the lens is immersed for 15 minutes only in a 0.25\%vol solution. These conditions are selected in order to promote the formation of dense monolayers without extensive polycondensation of the silanes. The silica lens is then rinsed and sonicated in toluene before drying. \\ 
In what follows, the PTES, APTES and OTS treated surfaces will be denoted as "propyl", "aminopropyl" and "octadecyl", respectively.\\
The contact angle of deionized water droplets on these surfaces were found to be 56, 62 and 104~$^o$ for the propyl, aminopropyl and octadecyl surfaces, respectively. The grafting of the silica lenses was assessed from ellipsometric microscopy performed on silanized silicon wafers which were grafted simultaneously with the silica lenses. The ellipsometric images of the silanized surfaces allow to measure an average thickness which combines the molecular size of the grafted molecules and their surface density, and the homegeneity of the silane layer. Both propyl and octadecyl surfaces are homogeneous over areas of the order of 1~\si{\square\milli\meter} and a pixel size of 2~$\mu$m$^2$ with variations below 0.1~\si{\nano\meter}. The average thickness is 0.4~\si{\nano\meter} and 2.3~\si{\nano\meter} for propyl and octadecyl respectivement, in agreement with the literature\cite{bureau_friction_2007, wang_growth_2003}. Accounting for the molecular size of the silanes, this shows that, compared to propyl, octadecyl is grafted more densily and exhibits longer chains extending outwards,  {consistently with the literature \cite{brzoska_silanization_1994}}.
For aminopropyl, the average thickness is 1.6~\si{\nano\meter}, greater than the monolayer thickness measured in the literature at 0.7~\si{\nano\meter} \cite{vandenberg_structure_1991}, and the layer is more heterogeneous with sparse localized defects about 5~\si{\nano\meter} in thickness and 10~\si{\micro\meter} wide which can be interpreted as multilayered islands. Since propyl and aminopropyl silanes have about the same molecular size, we conclude that aminopropyl layers are denser than propyl layers. Hence, the surface densities rank as follows: propyl < aminopropyl < octadecyl, consistently with the contact angle measurements which show hydrophobicity increases likewise.

As a reference, a non-functionalised silica lens cleaned with piranha solution and UV/ozone was also considered. The frictional properties of this freshly cleaned silica lenses were observed to change as a function of time as a result of progressive surface contamination (timescale $\sim$ hour). The friction results reported below corresponds to stabilized values of the lateral force. In comparison, the silanated glass lenses provided friction forces that were stable over time. \\

\subsection{Friction set-up}\label{subsec:setup}
\begin{figure} [!ht]
	\centering
	\includegraphics[width=1\linewidth]{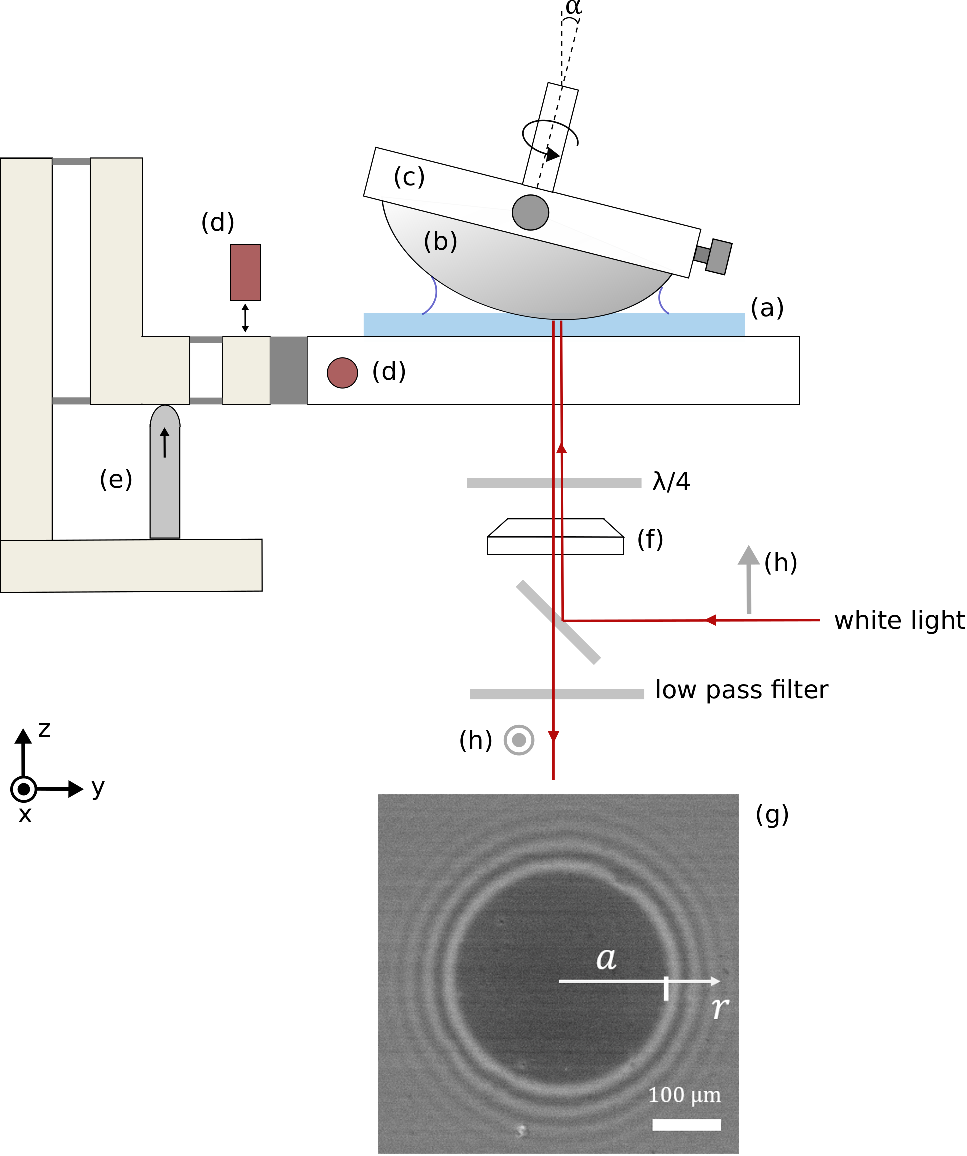}
	\caption{Schematic of the friction rotation set-up. The coated glass substrate (a) is contacting a silica lens (b) fixed on centering mount (c) which is actuated by a rotating stage. The rotation axis is inclined by an angle $\alpha=$5~deg. with respect to the $z$ axis by means of a goniometric stage. The normal and lateral force components ($F_n$ and $F_t$) are measured from the deflection of two crossed double-cantilever arrangements using two capacitive displacements sensors (d). A constant normal load is applied by means of a double cantilever beam system and a piezoelectric actuator (e) operated in feedback loop control. The contact immersed in a deionized water droplet is lightened by a white light source and images are recorded using a zoom lens (f), a CMOS camera (g) and an optical setup consisting in a semi-reflecting plate, two crossed polarizers (h) and a quarte-wave plate.}
	\label{fig:setup}
\end{figure}

A dedicated setup was designed in order to perform sliding experiments within a contact between a spherical silica lens and a gel film without any contribution of poroelastic transport to friction. Such a constraint requires that the contact remains fixed within the framework of the gel film. As schematically depicted in Fig.~\ref{fig:setup}, this goal was achieved by rotating the lens about an axis which is slightly tilted by an angle $\alpha=5$~deg. with respect to the normal to the film. For the considered radius of curvature of the lens ($R=23$~$\si{\milli\meter}$) and typical values of the contact radius $a \approx 100$~$\si{\micro\meter}$, geometrical calculations indicate that the radius of curvature of the sliding trajectories within the contact is about ten times the contact radius, \textit{i.e.} sliding trajectories can be considered as rectilinear within the contact. As compared to more conventional rotational experiments using plate rheometers in which sliding velocities vary from zero at the center to the set value at the edge, the advantage here is that the sliding velocity field within the contact is uniform to a very good approximation. \\
All the experiments are carried out under imposed normal load and sliding velocity with the contact immersed in a droplet of deionized water. As shown in Fig.~\ref{fig:setup}, an accurate alignment of the apex of the lens (b) with respect to the axis of the  rotating stage (M060-DG, PI) is obtained  by means of a home-made, two-axis, micro-metric translation stage (c) with high stiffness. During experiments, the normal and lateral forces are continuously monitored from the measurement (using capacitive displacement sensors CS02 and CS05, Microepsilon) of the deflection of two crossed double cantilever beams (d) with stiffnesses $k_{\perp}=1.01\:10^{5}$~N.m$^{-1}$ and $k_{\parallel}=1.13\:10^{5}$~N.m$^{-1}$ in the normal and lateral directions, respectively. A double cantilever beam actuated in servo-loop control by a piezoelectric actuator (e) allows to maintain a constant normal force during the rotation of the lens and to compensate the effects of small residual misalignments of the lens with respect to the rotation axis.\\
%
Reflection Interference Contrast Microscopy (RICM)~\cite{theodoly2010} images of the immersed contact were continuously recorded through the glass substrate using a CMOS camera (SVS Vistek Exo250 MU3, 2048x2048 pixels with 12 bits resolution), a combination of crossed-polarizers and a quarter-wave plate, a macro zoom (APO Z16, Leica), and white light illumination (HXP, Leica). A typical image of the steady-state contact is shown in Fig.~\ref{fig:setup}(g) where the contact appears as a dark disk whose radius is denoted $a$. The optical contrast measured at the gel/probe interface showed that no lubrication water film is trapped at this interface.\\
Friction experiments were carried out with sliding velocities ranging 
from 0.4~$\si{\micro\meter\per\second}$ to 450~~$\si{\micro\meter\per\second}$ and 
under imposed normal load $F_n$ from 20~$\si{\milli\newton}$ to 200~$\si{\milli\newton}$. The values of the normal load were chosen so that the hydrogel remains sufficiently hydrated\cite{delavoipiere_poroelastic_2016}. More precisely, the water volume fraction $\phi$ in the film is always larger than the threshold corresponding to the glass transition of the PDMA network ($\phi \approx 0.2$). In the range of loads tested, the contact radius $a$ varies between 100 and 150~$\mu$m so that it is always far larger than the film thickness $e_0$. Under such a mechanical  confinement, the polymer network compresses
along the normal direction solely with no significant lateral strain. As
a consequence, the number density of polymer chains in the
plane of the interface can be assumed to remain constant, independently of the level of film
compression.\\

In all experiments, sliding motion is initiated after a static contact time greater that the poroelastic time $\tau$ (of the order of a few tens of seconds for the contact conditions under consideration~\cite{ciapa_transient_2020}) in order to ensure that indentation equilibrium is achieved. However, when sliding starts, we systematically observe that the lens further sinks into the film: a small increase of the indentation depth $\delta$ (or equivalently of the contact radius $a$) occurs over a time close to the poroelastic time (results not shown). Typically, the increase in indentation depth is about 30~$\si{\nano\meter}$ for an initial indentation depth $\delta_0 \approx 410$~$\si{\nano\meter}$ under an imposed normal load $F_n=100$~$\si{\milli\newton}$ (the initial static contact radius $a_0=132$~$\si{\micro\meter}$ increases by 5~$\si{\micro\meter}$). It was carefully checked that this phenomenon is not an artifact due to a cross-talk between the normal and lateral directions. Instead, it reflects an intrinsic coupling between the normal and lateral stress components at the sliding interface which will be discussed at the end of this paper in Section \ref{sec:discussion}. In what follows, the average sliding stress $\sigma_t$ was systematically determined from the ratio of the steady-state friction force $F_t$ to the actual contact area $\pi a^2$ (see Fig.~\ref{fig:setup}g) measured under steady-sliding.\\
%
\section{Friction model}
We aim at discussing the variations of the friction stress with both interfacial physical-chemistry and velocity. For that purpose, we choose as a framework the friction model initially developed by  {Schallamach\cite{schallamach_theory_1963}} and revisited  {by Chernyak and Leonov\cite{chernyak_theory_1986, leonov_dependence_1990},} and Singh and Juvekar~\cite{singh_steady_2011} in which friction of elastomers is ascribed to a molecular mechanism, namely the elastic stretching and desorption of polymer chains at the sliding interface. In Section~\ref{sec:model}, we recall the main assumptions and results of the model. Then, in Section~\ref{sec:model:approx}, we identify various velocity regimes and we conveniently derive approximate analytical expressions for the frictional shear stress versus velocity, with parameters related to the molecular interactions and to the stretching behavior of polymer chains. We especially identify an intermediate velocity range in which $\sigma_t$ depends logarithmically on sliding velocity.\\
\subsection{Steady-state frictional stress}\label{sec:model}
\begin{figure}[!ht]
  \centering
  \includegraphics[width=1\linewidth]{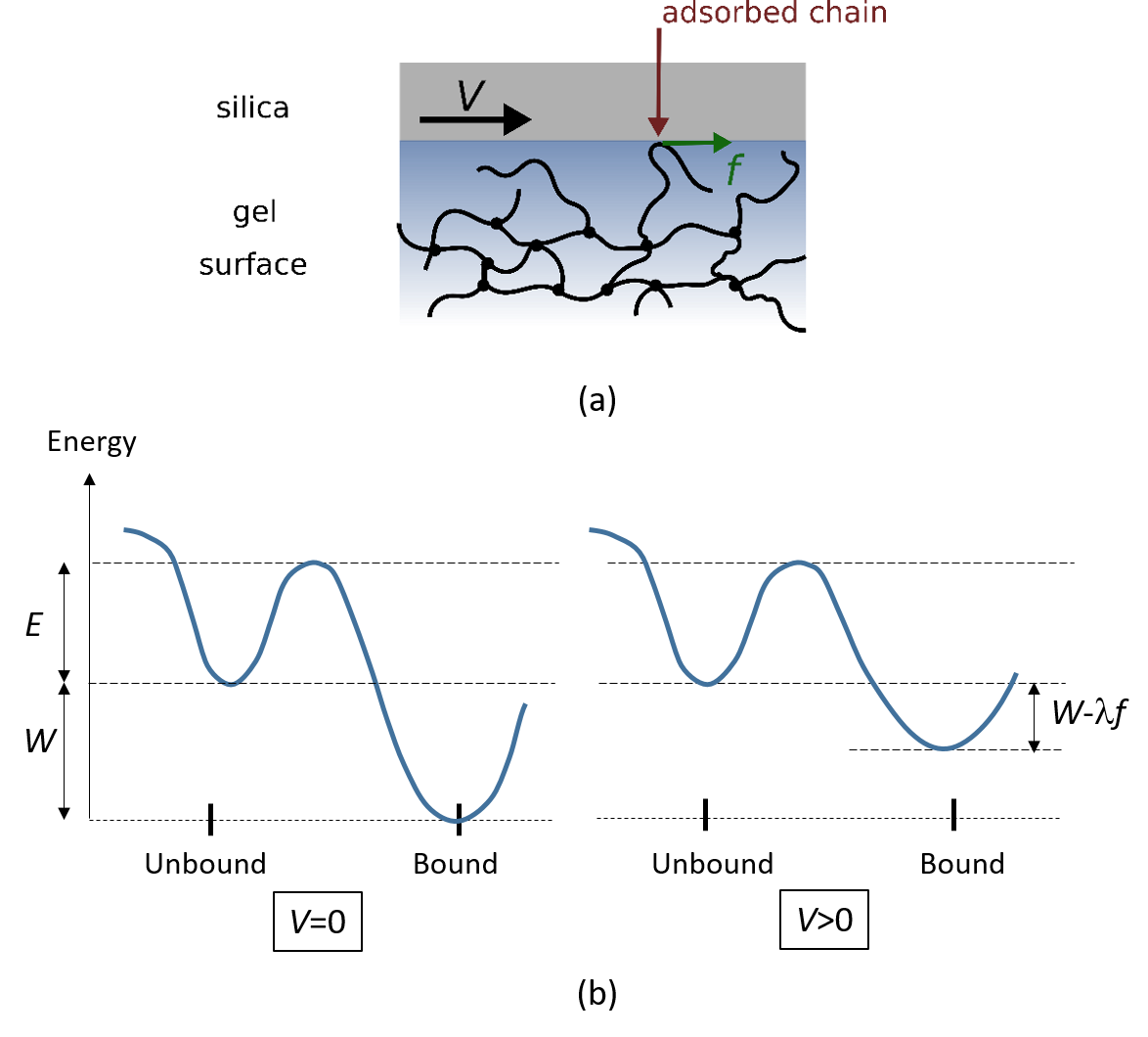}
\caption{Schematics of the Schallamach's model: (a) The silica surface slides at a velocity $V$ on the gel surface; if a chain from the polymer network adsorbs on silica, it stretches and experiences a force $f$ assumed to be oriented along the interface plane; (b) Energy profiles for adsorption and desorption of the polymer chains at $V=0$ and $V>O$. The energy difference between bound and unbound states at rest is $W$. Upon sliding, this energy is biased by an amount $\lambda f$ where $\lambda$ is an activation length. }
	\label{fig:schallamach}
\end{figure}

In the following, an infinite extended  gel/solid interface is considered. At rest, the adsorption of the polymer chains in the gel is thermally favored and reversible: the adsorption and desorption rates are described by probability distributions of Arrhenius' type where the energy barriers are denoted $E$ and $E+W$ respectively, as depicted in Fig.~\ref{fig:schallamach}: Here, $W$ is a molecular adhesion energy and we define $u=e^{-W / k T}$. Hence, at rest, 
the lifetimes of an unbound chain $\tau$ and of a bound chain $\tau_{ads}$ can be written as:
\begin{align}
\tau&=\tau_{0} e^{E / k T} \\ 
\tau_{ads}&=\frac{\tau}{u}=\tau_{0} e^{(E+W) / k T}
\label{eq:tau_ads_def}
\end{align}
%
where $\tau_0$ is a molecular characteristic time of exploration often taken as $h/kT$, with $h$ the Planck's constant. 
Upon sliding at a velocity $V$, the adsorbed chains are stretched and the energy landscape is biased accordingly. If a chain remains attached for a time $t_a$, it experiences a force $f$ that increases with $t_a$ and favors desorption. Then, following Eyring's model, the bias can be written as the stretch energy stored in the bond $\lambda f$ where $\lambda$ is an activation length. For polymeric gels, this length is expected to be of the order of the gel mesh size. Here, we assume the chains are stretched within the plane of the interface so that the force $f$ is directed along the interface plane as depicted in Fig.~\ref{fig:schallamach}a). Hence, denoting the chain length at rest $L$, and the stretch, $\Delta L$, the stretch varies linearly with the bond time and the velocity as $\Delta L=V t_a$.
In the present derivation of the model, assumption is made that the unbound to bound transition is not biased by stretching\footnote{The effect of stretching on the unbound to bound transition is cautiously accounted for in models derived for the detachment force of adhesive patches and is found to affect the very low velocity regime. See for example \cite{friddle_interpreting_2012}.} while the unbinding rate depends on the lifetime of the bond $t_a$ through $f(t_a)$. 
 {In addition, $t_a$ is the age of the bond, but we assume no chemical ageing : the energies $E$ and $W$ do not depend on $t_a$.} Then, during sliding, the binding and unbinding rates per unit time follow Eyring's model\cite{eyring_viscosity_1936} and are written as:
\begin{align}
 r_b&=\frac{1}{\tau_0}e^{-E/kT}=\frac{1}{\tau}\\ 
 r_u&=\frac{1}{\tau_0}e^{-\left( E+W-\lambda f\left( t_a \right) \right)/kT}=\frac{u}{\tau}e^{-\lambda f\left( t_a \right)/kT}
 \label{eq:rbru}
\end{align}
We now derive the equations governing the distribution $n(t,t_a)$ that describes the number of bonds  {that have been attached for a time} $t_a$ at a time $t$.  Indeed, this distribution allows to express the friction stress $\sigma_t$, which is  the sum of the forces exerted by the stretched chains at a given time $t$, according to: 
\begin{equation}
 \sigma_t=\int_0^\infty n\left( t_a,t \right)f\left( t_a \right)dt_a
\label{eq:sigmat}
\end{equation} 
We first define the total number of bonds $N(t)$ existing at time $t$ and given by:
\begin{equation}
 N(t)=\int_0^\infty n\left( t_a,t \right)dt_a
\label{eq:N}
\end{equation} 
Between $t$ and $t+dt$, the number of bonds aged $t_a$ increases, first, because of the ageing of the bonds that were aged $t_a-dt$ at time $t$. Second, old enough bonds unbind with a rate $r_u$ and in proportion to the number of bonds. Third, new bonds are created with a rate $r_b$, in proportion to the total number of available sites $(N_0-N(t))$, and are aged $t_a=0$ so that we use the Dirac function $\delta(t_a)$. Using Eqs.~\ref{eq:rbru}, the time evolution of the number of bonds obeys:
\begin{equation*}
 \frac{\partial n\left( t_a,t \right)}{\partial t}=
 -\frac{\partial n\left( t_a,t \right)}{\partial t_a}
 +\frac{1}{\tau}\left[N_0-N\left( t \right) \right]\delta\left( t_a \right) 
 -\frac{u}{\tau}e^{\lambda f\left( t_a \right) /kT}n\left( t_a,t \right)
\end{equation*} 
In steady state, the time derivative vanishes so that:
\begin{equation}
 \frac{dn\left( t_a \right)}{dt_a}
 =\frac{1}{\tau}\left(N_0-N \right)\delta\left( t_a \right) 
 -\frac{u}{\tau}e^{\lambda f\left( t_a \right) /kT}n\left( t_a \right)
 \label{eq:diff:nta}
\end{equation} 
This differential equation is solved by setting: $n(t_a)=C(t_a)g(t_a)$ where $g(t_a)$ is defined by:
\begin{equation}
g(t_a)=\exp\left(-\frac{u}{\tau}\int_0^{t_a}e^{\lambda f(\xi)/kT}d \xi\right)
\label{eq:gta}
\end{equation}
Equation~\ref{eq:diff:nta} becomes $C'(t_a)=\frac{1}{\tau}\left(N_0-N\right)\frac{\delta(t_a)}{g(t_a)}$ which yields: 
\begin{equation}
     C(t_a)=\frac{1}{\tau}\left(N_0-N\right) ~\text{for}~ t_a>0
\end{equation}
Then, the number distribution of bonds aged $t_a$ is given by:
\begin{equation}
    n(t_a)=\frac{1}{\tau}\left(N_0-N\right)g(t_a)
\end{equation}
We define $G$ as $G=\frac{1}{\tau}\int_0^\infty g(t_a)dt_a$. Equation~\ref{eq:N} gives the total number of active bonds $N$ and, from this, the distribution $n(t_a)$:
\begin{align}
    N=\frac{G}{1+G}N_0\\
    n(t_a)=\frac{\frac{1}{\tau}g(t_a)}{1+G}N_0
\end{align}
The friction stress can then be expressed from Eq.~\ref{eq:sigmat} as:
\begin{equation}
    \sigma_t=\frac{\frac{1}{\tau}\int_0^\infty g(t_a)f(t_a)dt_a}{1+G}N_0=\frac{N}{G}\frac{1}{\tau}\int_0^\infty g(t_a)f(t_a)\,dt_a
\end{equation}
which corresponds to Eq.~17 in reference~\cite{singh_steady_2011}. In the particular case of a  {Hookean} force $f$ depending linearly of the stretch $\Delta L=t_a V$ though a stiffness denoted $M$, the force $f(t_a)=MVt_a$ and $g(t_a)$, $G$ and $N$ can be further derived. The hookean model well describes the case of polymer chains in a poor solvent\cite{rubinstein_polymer_2003} as PDMA in water for which the Flory parameter is $\chi=0.57$ \cite{delavoipiere_swelling_2018}. 
A characteristic velocity $V^*$ comes up, for which the typical elastic energy stored in the stretched polymer chain $MV^*\tau\lambda$ is equal to the thermal energy $kT$. It corresponds to a characteristic stretch denoted $l=\frac{kT}{\lambda M}$. This allows to define two non-dimensional variables in time $s$ and space $z$ so that:
\begin{align}
    V^*=\frac{l}{\tau}=\frac{kT}{\tau\lambda M}\\
    t_a=\tau\frac{V^*}{V}s\\
    \xi=\tau\frac{V^*}{V}z
\end{align}
and Eqs.~\ref{eq:gta} and $G(t_a)$ become:
\begin{align}
    g(t_a)=\exp{\left(-\frac{uV^*}{V}\int_0^se^z\,dz\right)}=\exp{-\frac{uV^*}{V}\left(e^s-1\right)}\\
    G=\frac{V^*}{V}\int_0^\infty\exp{\left(-\frac{uV^*}{V}(e^s-1)\right)} \,ds
    \label{eq:G:complet}
\end{align}
The velocity dependence of the number of bonds $N=N_0\frac{G}{1+G}$ and of the friction stress
\begin{equation}
    \sigma_t=\frac{kT}{\lambda}\frac{V^*}{V}\frac{N}{G}\int_0^\infty \exp{\left(-\frac{uV^*}{V}(e^s-1)\right)} s\,ds
    \label{eq:sigma:complet}
\end{equation}
are set by two parameters, namely $V^*/V$ and $uV^*/V$. The former 
is actually the ratio between the {\it advective} time $\tau_V=l/V$ built upon the characteristic stretch $l=\frac{kT}{\lambda M}$ and the lifetime of a {\it detached} bond $\tau$. The latter is a ratio between this advective time $\tau_V=l/V$ and the lifetime of an {\it active} bond $\tau_{ads}$ defined in Eq.~\ref{eq:tau_ads_def}.  Eq.~\ref{eq:sigma:complet} is evaluated in Appendix, and the result is plotted in Fig.~\ref{Fig:singh}. A bell shaped curve is obtained for $\sigma_t(V/V^*)$ with a maximum at $V=V^*$. 
\begin{figure}[!ht]
	\centering
	\includegraphics[width=1\linewidth]{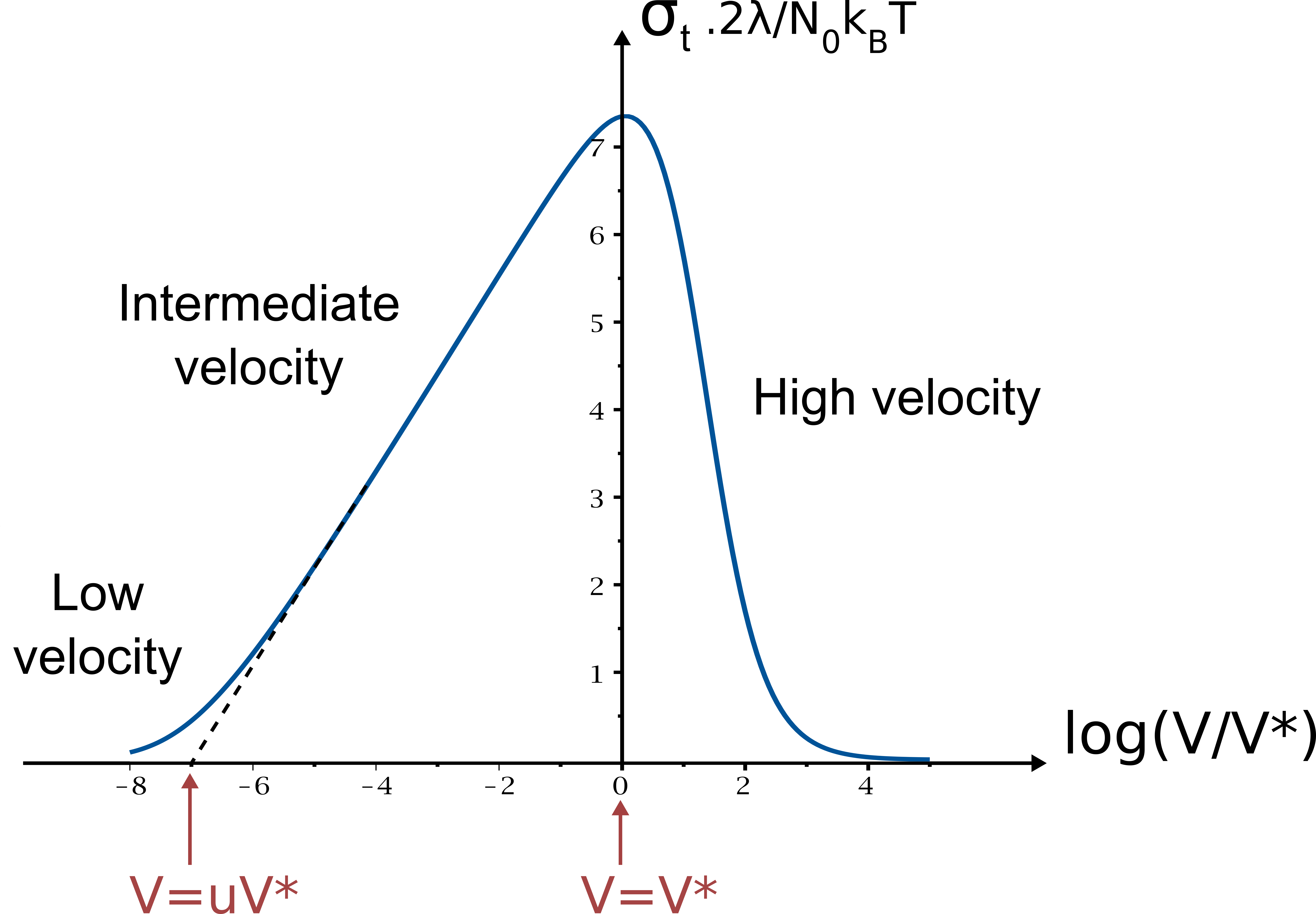}
	\caption{Friction stress $\sigma_t$ normalized by $N_0kT/2\lambda$ as a function of the normalized velocity $\hat{V}=V/V^*$ in logarithmic scale (Eq.~\ref{eq:sigma:complet}). Here, $u=10^{-7}$ which corresponds to an adhesion energy $W\sim 15 kT$. Intermediate velocities correspond to the logarithmic regime.}
	\label{Fig:singh}
\end{figure}
\subsection{Approximate expressions for $\sigma_t(V)$}\label{sec:model:approx}
In the following, we offer to derive approximate analytical expressions of $\sigma_t$ at low, intermediate, and large velocities.\\

\subsubsection{Low velocity regime $V\ll uV^*$}
At low velocities, the integrand in Eq.~\ref{eq:G:complet} almost always vanishes except for vanishing $s$ values. Hence, Watson's lemma yields $G\sim \frac{V^*}{V}\int_0^\infty \exp\left(-s\frac{uV^*}{V}\right)\,ds\sim\frac{1}{u}$. Remember that $u$ is defined as $u=e^{-W/kT}$ where $W$ is a molecular adsorption energy which is typically equal to a few $kT$ \cite{rubinstein_polymer_2003}. Hence, $u\ll1$ and, from this, $G>>1$. It follows that the total number of bonds $N\sim N_0$: almost all possible sites are bound. Here, the sliding is slow enough so that the advective time is large compared to both the adsorption and desorption times. It follows that the friction stress increases linearly with the velocity as: $\sigma_t=\frac{N_0kT}{\lambda}\frac{V}{uV^*}=N_0MV\tau_{ads}$.\\
Experimentally, this low velocity regime is not observed for all the silica lenses and will not be further examined.\\

\subsubsection{Intermediate velocities $uV^*\ll V\ll V^*$}
At larger velocities for which $uV^*\ll V$, the integrand in Eq.~\ref{eq:G:complet} is a step-down function: its value is zero for large enough $s$ and 1 below a threshold value $S$ given by: $\exp\left(-\frac{uV^*}{V}(e^S-1)\right)=1/2$ so that $S=\ln \frac{V \ln 2}{u V^*}$. It follows that the integral in $G$ and $\sigma_t$ is simply approximated as:
\begin{equation}
    G=\frac{V^*}{V}\ln \frac{V \ln 2}{uV^*}
    \label{eq:approx:large_velocities:1}
    \end{equation}
    \begin{equation}
    \sigma_t=\frac{N_0kT}{2\lambda}\frac{S^2}{\frac{V}{V^*}+S}
    \label{eq:approx:large_velocities:2}
\end{equation}
This allows to define two velocity regimes, above and below $V^*$.\\
Below $V^*$, this intermediate regime is clearly logarithmic in velocity. We find $G>>1$ and again, the total number of bonds is almost equal to the number of possible sites $N\sim N_0$: the thermal energy controls the binding and unbinding transitions and the bias on the energetic landscape due to the elastic force $f$ is negligible. Hence, the mean lifetime of a bond is still equal to $\tau_{ads}$ but this time, the advective time is smaller than the lifetime of a bond at rest: $\tau\ll\tau_V\ll\tau_{ads}$ so that the sliding effectively stretches the chains and, during the adsorption time, the elastic stretch increases with velocity. The friction stress increases faster with velocity, and is now written as:
\begin{equation}
    \sigma_t=\frac{N_0kT}{2\lambda}\ln \frac{V \ln2}{uV^*}
    \label{eq:siglogap}
\end{equation}

\subsubsection{High velocities $uV^*\ll V^*\ll V$}
Finally, above $V^*$, Eqs.~\ref{eq:approx:large_velocities:1},\ref{eq:approx:large_velocities:2}  still hold but, now, the large velocity regime corresponds to a decrease in the number of bonds due to the detrimental effect of sliding on the binding transition: here, $\tau_V\ll\tau\ll\tau_{ads}$. It follows from Eq.~\ref{eq:approx:large_velocities:1} and $V\gg>V^*$ that $G\ll1$ and $N\sim N_0 G \ll N_0$: the number of active bonds decreases. In the meantime, the stretching also diminishes the mean bond time which decreases with $V$. Altogether, the friction stress decreases with velocity: $\sigma_t\sim\frac{N_0kT}{2\lambda}\frac{V^*}{V}\ln^2\frac{V \ln 2}{uV^*}$. In this case, this equation only partially approximates the exact solution since the active bonds number decreases slowly. In the literature, this decreasing and large velocity regime was positively compared \cite{singh_steady_2011} to data collected in friction experiments of silicone elastomers on alkyl-silanized glass slides \cite{vorvolakos_effects_2003}. Qualitative comparisons were also made with friction experiments on hydrogels of varied chemistry where bell-shaped curves were observed \cite{gong_gel_1998}.\\

\section{Experimental results}
\begin{figure} [!ht]
	\centering
	\includegraphics[width=1\linewidth]{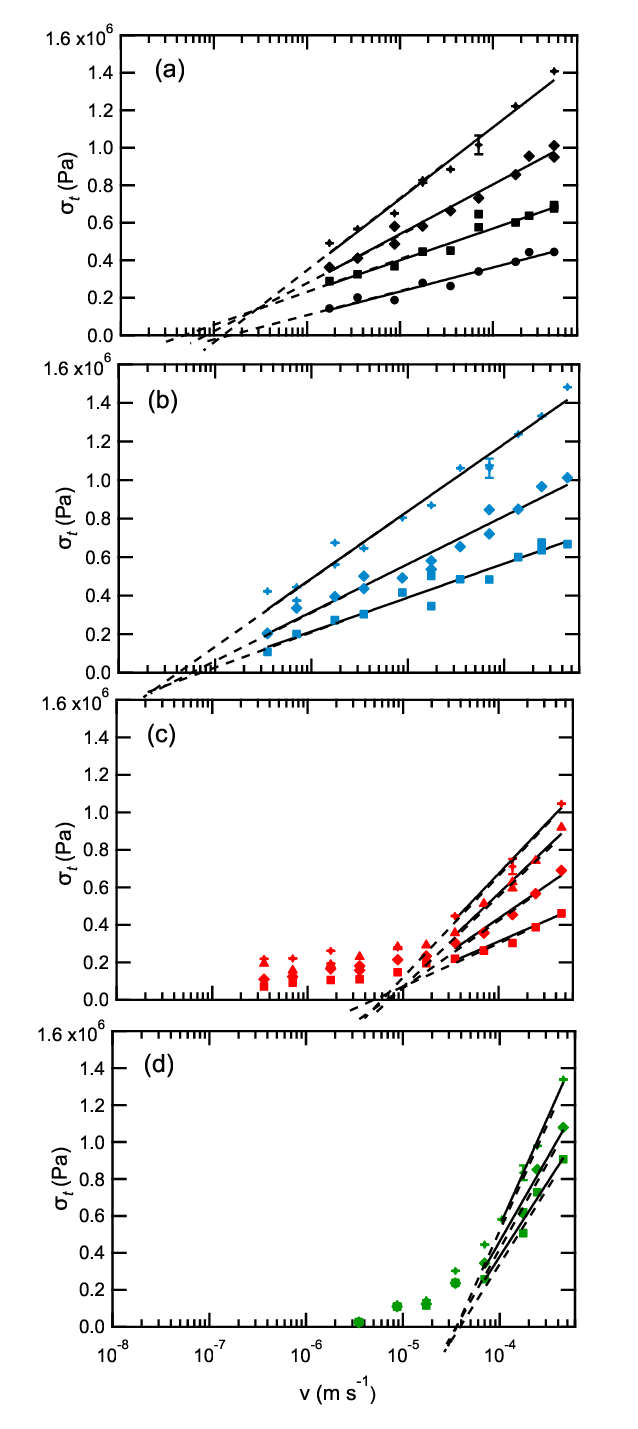}
	\caption{Friction stress $\sigma_t$ as a function of the sliding velocity $v$. Film thickness $e_0$=2.3~$\mu$m, swelling ratio $Sw=1.8$. Surface treatment of the silica lens: (a) bare silica; (b) propyl; (c) aminopropyl; (d) octadecyl. Normal load $F_n$: ($\bullet$) 20~\si{\milli\newton}; ($\blacksquare$) 50~\si{\milli\newton}; ($\blacklozenge$) 100~\si{\milli\newton}; ($\blacktriangle$)  150~\si{\milli\newton}; (\FourStar) 200~\si{\milli\newton}. Solid and dotted lines are fits to Eq.~\ref{eq:sigmafitlogV}.}
	\label{fig:sigmat_v}
\end{figure}

For each of the functionalised lenses, the steady state friction force $F_t$ and contact radius $a$ were measured as a function of the sliding velocity $V$. From this raw data (detailed in Supplementary Information),  the average frictional shear stress $\sigma_t=F_t/\pi a^2$ was calculated and reported in Fig.~\ref{fig:sigmat_v} as a function of the sliding velocity $V$ plotted on a logarithmic scale, for the film with thickness $e_0$=2.3~$\mu$m and $Sw=1.8$  for varied normal loads $F_n$. The four data sets clearly exhibit logarithmically increasing branches. Both their slope and intercept with the $x$-axis strongly depend on the physical-chemistry of the silica surface. The slope also depends on the normal load. 
Accordingly, we offer to analyse our data in the light of the model derived for intermediate velocities (Eq.~\ref{eq:siglogap}) for which the thermal energy controls the transitions between free and bound states and stretching energy shortens the unbinding.\\ 
%

\begin{figure} [!ht]
	\centering
	\includegraphics[width=1\linewidth]{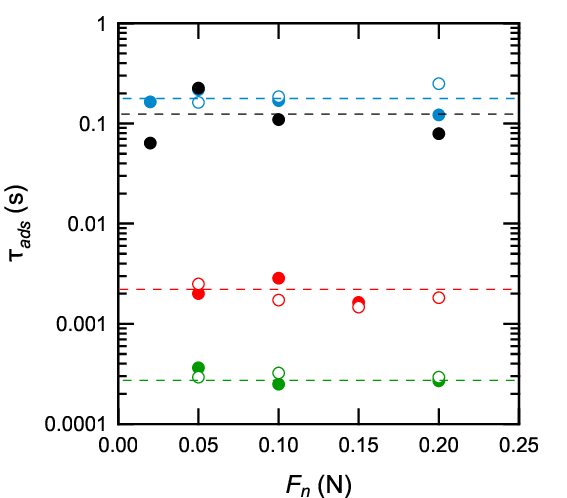}
	\caption{Lifetime in the bound state $\tau_{ads}$ as a function of normal load $F_N$. $\bullet$ bare silica, {\textcolor{cyan}\textbullet} propyl, {\textcolor{red}\textbullet}  aminopropyl, {\textcolor{green}\textbullet}  octadecyl. Open and filled symbols refer to films with $e_0=2.8$~$\si{\micro\meter}$, $\lambda=2.3$ and $e_0=2.3$~$\si{\micro\meter}$, $\lambda=1.8$, respectively. Dotted lines are guides for the eye.}
	\label{fig:tau_ads}
\end{figure}

\begin{figure} [!ht]
	\centering
	\includegraphics[width=1\linewidth]{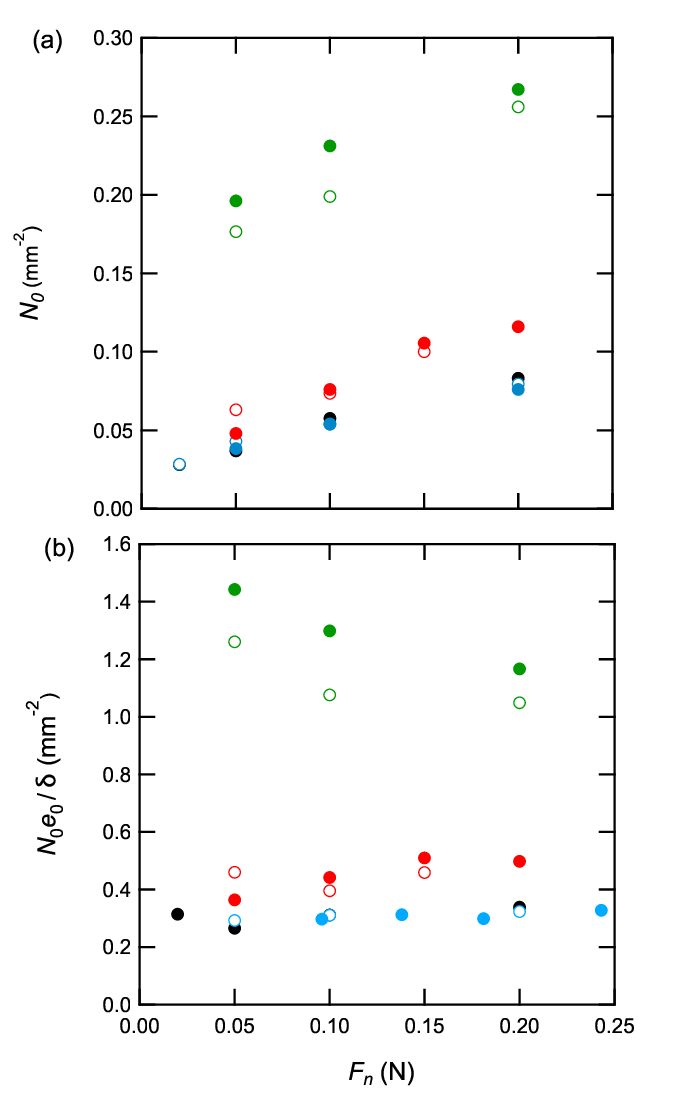}
	\caption{ Density of available binding sites on the silica surface as a function of the normal load $F_N$. (a) Number of binding sites per surface area $N_0$; (b) Reduced number of binding sites $\tilde{N_0}=N_0 e_0 / \delta$. $\bullet$ bare silica, {\textcolor{cyan}\textbullet} propyl, {\textcolor{red}\textbullet}  aminopropyl, {\textcolor{green}\textbullet}  octadecyl. Open and filled symbols refer to films with $e_0=2.8$~$\si{\micro\meter}$, $\lambda=2.3$ and $e_0=2.3$~$\si{\micro\meter}$, $\lambda=1.8$, respectively.}
	\label{fig:N0_N0red}
\end{figure}

Since the two gel films under consideration have similar swelling ratios, the order of magnitude of all the parameters related to the polymeric network will be kept as a constant. 
 {Noticeably, the stiffness $M$ of the bonds can here be precisely modelled from the stiffness of the network chains, at variance with studies on patches of self-assembled monolayers (SAM) where the bond stiffness could not be separated from the bonds surface density.}\cite{drummond_friction_2003} 
The chain stiffness $M$ is estimated as  {$M=kT/\nu_c b^2$} with  {$b\approx 1$~nm the Kuhn length for PDMA and $\nu_c=10$ the number of segments of a chain at the interface, which we consider equal to the segment number between crosslinks measured in Section~\ref{sec:hydrogel}}. 
Hence, the characteristic stretching length $l=\frac{kT}{\lambda M}$ can be rewritten as $l\sim \frac{\nu_cb^2}{\lambda}$, with $\lambda \sim 1$~nm (the mesh size): we find $l\sim 10$~nm.  {This value is 20 times larger than the stretch obtained on SAMs of short dodecyl chains, of 0.45~nm.\cite{drummond_friction_2003} This relatively large value reflects the compliance of the coiled polymer chains in the present hydrogel system, which conveniently shifts the velocity range for molecular friction towards larger values since $V^*=l/\tau$.}
\\
On the other hand, the lifetime of a bond will depend on the physical chemistry of the gel/silica interface though $\tau_{ads}=\tau/u$, the lifetime of a bond at zero velocity, defined by Eq.~\ref{eq:tau_ads_def}. Hence, omitting the $\ln{2}\sim1$ factor, Eq.~\ref{eq:siglogap} is conveniently rewritten as:
\begin{equation}
\sigma_{t}=N_{0} \frac{k T}{2 \lambda} \ln \frac{V}{l/\tau_{ads}}
\label{eq:sigmafitlogV}
\end{equation}
 {in which $l$ depends on the polymeric network only, while $N_0$ and $\tau_{ads}$ characterize the interfacial adsorption.} Equation~\ref{eq:sigmafitlogV} was fitted to the logarithmic branch of the friction stress data from Fig.~\ref{fig:sigmat_v} with the adsorption time  $\tau_{ads}$ and the surface density of bonds $N_0$ as fitting parameters. For propyl and bare silica (Fig.~\ref{fig:sigmat_v}a,b), the logarithmic branch extends over more than three decades in velocity, and over one decade for aminopropyl and octadecyl (Fig.~\ref{fig:sigmat_v}c,d). The fitting parameters $\tau_{ads}$ and $N_0$ are plotted in Fig.~\ref{fig:tau_ads} and~\ref{fig:N0_N0red}a as a function of the normal force $F_n$ for the four different silica chemistries (colors) and the two PDMA films (empty or full markers). \\
Remarkably, for each type of grafted silica, the stress-velocity curves intercept the $x$-axis at the same value of the critical velocity $uV^*$ independently of the normal force or of the hydrogel films. It follows that each silica treated lens is characterized by a single value of $\tau_{ads}$. For propyl silica, the mean value is $\tau_{ads}=0.2$~s. Noting for example that the propyl curve exhibits a logarithmic branch over 4 decades and no maximum in the experimental window allows to estimate that $u$ exceed $10^{-4}$ so that the molecular adhesion energy $W$ exceeds $4kT$.\\
%
On the other hand, the number of available sites $N_0$ is found in Fig.~\ref{fig:N0_N0red}a to increase with $F_n$, a surprising result which suggests a coupling between the frictional and normal stress responses. It nevertheless reduces to a linear dependence on the elastic deformation of the film which is squeezed by the lens: when $N_0$ is normalized by the ratio between the actual indentation depth of the lens into the film $\delta$ and the swollen thickness of the hydrogel film $e_0$, the data for the two different gel films no longer varies with the normal load as plotted in Fig.~\ref{fig:N0_N0red}b. This result allows to compare the different types of surface in terms of number density of available sites: bare silica, propyl, and aminopropyl silica engage about the same number of bonds with the hydrogel, while for octadecyl silica, this number is 3 to 4 times larger:  $N_{0,propyl} \sim N_{0,silica}\lesssim N_{0,amino}\ll N_{0,octadecyl}$. We will see that this ranking is reversed when adhesion energy is considered.\\
%
In the following, we offer to compare the adsorption times depending on the silica chemistry. The largest value of $\tau_{ads}$ being obtained for bare silica and propyl, these interfaces exhibit the largest adhesion energy and a ranking in terms of adhesion is established:
$W_{propyl} \sim W_{silica} > W_{amino} > W_{octadecyl}$ with an estimate of the differences: $W_{propyl} - W_{silica} \sim \pm 0.5  kT$, $W_{propyl}-W_{amino}\sim 4.6  kT$ and $W_{propyl}-W_{octadecyl}\sim 6.5   kT$. 
\section{Discussion}\label{sec:discussion}

\begin{figure} [!ht]
	\centering
	\includegraphics[width=0.8\linewidth]{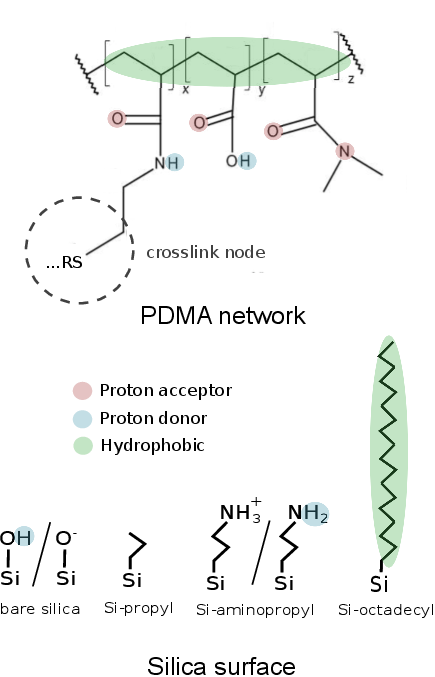}
	\caption{Schematics of the available groups for binding at the surface of the PDMA network (top) and of the functionalised silica surfaces (bottom). The proton acceptors and donors involved in hydrogen bonds are indicated together with the groups possibly involved in entropic (or hydrophobic) interactions.}
	\label{fig:interactions_schematic}
\end{figure}

The interface of PDMA with bare silica and propyl silica having close adhesion energy are likely to engage molecular interactions of the same type, namely H-bonds between amides groups of PDMA and silanols \ce{SiOH} on the silica side (see Fig.~\ref{fig:interactions_schematic}). Indeed, the silanization of silica by short alkyl chains is only partial, disordered and unreacted silanol groups are still available as compared to longer alkyl chains  {as detailed in Section~\ref{sec:silica:charac}}. Here, the grafted propyl chains appear to be short enough not to disturb the H-bonds with neighboring silanols. This is at variance with what is observed with the long octadecyl C18 tails: H-bonds with the buried silanol groups are probably unfavored. Instead, an entropic attraction is likely to develop between the densely grafted layer of C18 tails and the PDMA chains which excludes water. It is also thought of as an "hydrophobic interaction" and expected to correspond to much lower adhesion energies than H-bonds.\cite{kekicheff_long-range_2019}  {Recalling from Section~\ref{sec:silica:charac}} that the silanization by OTS provides both larger and more uniform grafting densities compared to propylsilane,\cite{wang_growth_2003, brzoska_silanization_1994} 
 the surface density $N_{0, octadecyl}$ we measure is consistently larger than $N_{0, propyl}$. Finally, the aminopropyl-grafted surface has an intermediate behavior. The tails of the aminopropyl and propyl silanes have the same length, 3 carbons. In water at pH=6, part of the amine groups of $pKa\simeq9.5$ \cite{shen_effect_2019} are protonized into  \ce{NH3+} while part of the silanol are deprotonized into \ce{SiO-}. Electrostatic interactions between neighboring charged aminopropyl silanes and silanols lead to the bending of the former on the latter, preventing the formation of H-bonds with part of the silanols which are then capped by hydrophobic C3 tails. Hence, on the aminopropyl-silica surface, the average adhesion energy with PDMA in water results from a combination of H-bonds with the remaining non-capped silanols and low hydrophobic interactions elsewhere: this could explain the low value of $W_{0,amino}$ as compared to bare silica or propyl silica.\\
%

We shall now briefly comment on the two observations evidencing a coupling between the normal and friction forces: {\it (i)} Upon sliding, the silica lens further sinks into the hydrogel film, whatever its surface chemistry. The contact radius increases. (Section~\ref{subsec:setup}) {\it (ii)} The prefactor $N_0$ of the logarithmic $V$-dependence of the friction stress (Eq.~\ref{eq:sigmafitlogV}) increases with the normal force and linearly depends on the normal deformation of the gel (Fig.~\ref{fig:N0_N0red}). A complete description of these coupling mechanisms would deserve a comprehensive study, but we suggest here to link these observations with the theoretical descriptions derived by Leonov \cite{leonov_dependence_1990, chernyak_theory_1986} of the adhesive friction of elastomers. Based on  {Schallamach's model}, Leonov accounts for the finite non-zero thickness of the interface whereas both Singh,  {Schallamach} and, from them, the present model, assume it to be zero. Then, if the polymer chains are attached to the gel network at a certain non-zero depth from the silica interface, upon sliding and stretching, they are likely to exert an elastic force $f$ with a component both along the interface and normal to it: Leonov shows that this mechanism results in not only a friction stress but also a normal stress: the gel pulls the interface in with a normal stress that increases by up to 10$\%$ when the velocity is ten-fold. The derivation of a model combining molecular adsorption, stretching, and desorption {\it à la Singh} to an interface with finite non-zero thickness {\it à la Leonov}, and comparison with our experimental data on normal/friction coupling will be the subject of a forthcoming paper.\\
 {On the experimental side in the literature, friction was found to depend on the load when measured between two monolayers of alkyl chains\cite{drummond_friction_2003}, but the load-dependence was ascribed to an increase of the adsorption energy or, equivalently, of the bond lifetime, at variance with the present findings, with no explanation of the underlying mechanisms.} 
{We offer next to summarize our results by expressing the friction force as a function of the contact radius, and extend their validity to another geometry where the silica lens slides on the gel film following a linear trajectory. The rectilinear set-up we used was described in earlier works \cite{delavoipiere_friction_2018, ciapa_transient_2020} that evidenced a poroelastic flow within the gel film in response to the lens displacement. First, using the geometrical relationship between indentation depth $\delta$ and contact size $a$, $\delta=a^2/R$, our experimental data show that $\sigma_t$ vary with the contact radius as $a^2$ or equivalently, $F_t$ varies as :
\begin{equation}
F_t=a^4 \tilde{N_0}\frac{kT}{2e_0\lambda}  \ln \frac{V}{l/\tau_{ads}}
\label{eq:Ft:a:final}
\end{equation}
with $\tilde{N_o}$ the reduced number of bonding sites per surface area that only depends on physical-chemistry.
As detailed in Supplementary Material, for both geometries, namely rotational and rectilinear friction, we find that (i) rectilinear friction depends on surface chemistry, (ii) a master curve in $F_t/a^4$ allows to collapse the data as expected from Eq.~\ref{eq:Ft:a:final}. These findings show that (i) friction by molecular adsorption at the interface occurs even when poroelastic flows come into play, and (ii) even dominates over poroelastic dissipation. Poroelasticity, however, does set the contact size in transient sliding and steady state rectilinear sliding \cite{delavoipiere_friction_2018,ciapa_transient_2020}.}

Finally, we briefly discuss the friction data in the very low velocity regime which is not accounted for by the present model (Fig.\ref{fig:sigmat_v}c,d and SI2), in which the effect of stretching on the unbound to bound energy barrier is neglected. From the literature on pull-off experiments on small adhesive patches\cite{friddle_near-equilibrium_2008}, this hypothesis, however, is expected to fail at very low velocity where the pinning energy barrier should be also be biased. This remark calls for the development of a model combining biased interaction potentials as in \citet{friddle_near-equilibrium_2008} and variable number of available sites as in here\cite{chernyak_theory_1986, drummond_friction_2003, singh_steady_2011}. We also note that ageing effects are likely to exist in this low velocity regime where sites on the functionalised silica surface spend longer times within the contact\cite{singh_model_2021}.
%
\section{Conclusions}

Steady-state friction experiments were conducted on hydrogel thin films using a combination of model interfaces with controlled and varied surface chemistry and an original experimental set-up. The latter relies on the sliding of a rotating solid sphere at the hydrogel surface in a geometrical arrangement that allows to (i) obtain a uniform sliding velocity within the sliding interface, (ii) reach the low velocity regime in which interfacial friction dominates, (iii) provide direct and contrasted images of the gel/probe interface, (iv) provide a fine measurement, over time, of the friction and normal forces, and of the contact area, indentation depth, and possibly lubrication layers. In this geometry, by working at the bare surface of model hydrogel films sliding against functionalised silica surfaces, we show that friction is dominated by adhesive molecular interactions resulting in irreversible adsoprtion-stretching-desorption mechanisms and thus dissipation at the sliding interface.\\
An analytical model built upon the work of \citet{schallamach_theory_1963} reformulated by \citet{singh_steady_2011} allowed us to derive analytical expressions for the friction stress as a function of velocity. A logarithmic branch was clearly identified in both the model and the experimental data. By fitting the friction stress versus velocity data to the model for the four surface chemistries tested and under varied load, we successfully assess the molecular parameters involved in an adsorption/stretching/desorption mechanism of the polymer chains of the hydrogel network at the solid interface.  {We insist that this was made possible notably because the stiffness of the bonds is set by the elasticity of the polymeric chains independently of the counter-surface physico-chemistry, and the relatively dilute surface of hydrogels allows to hypothetise that bonds are independent, two conditions underlying the model that are not easily met in experimental systems.\cite{drummond_friction_2003, bureau_friction_2007}}

Both the adsorption energies and the surface densities of bonds we measure are consistent with the physico-chemical characteristics of the functionalised silica surfaces. Hence, for the first time, the dissipative effects of molecular interactions at a sliding hydrogel interface were isolated from other sources of dissipation involved in friction (no viscous dissipation, no poroelastic flow). \\
We believe our work has strong implications from both applied and fundamental points of view. In material science, when frictional properties of hydrogels are to be controlled, our results provide a quantitative tool to assess the most relevant parameters to be tuned. In soft matter physics, friction on hydrogel films in a rotating sliding geometry appears as a model situation to explore the interaction energy landscapes at a polymer/solid interface.   
\section*{Acknowledgements}
The authors gratefully acknowledge the support of Jean-Claude Mancer in the realization of the frictional set-up. We also thank R. Bennewitz and C. Drummond for fruitful discussions and for kindly making us aware of the work of Friddle and coworkers.\\
%
\appendix
\section*{Appendix}
\numberwithin{equation}{subsection}
\renewcommand{\thesubsection}{\Alph{subsection}}
\subsection{Explicit evaluation of $\sigma_t(V)$}
We derive here the explicit calculation of the friction stress, starting from the equations for $g(t_a)$, $G$, and $\sigma_t$, Eq.~\ref{eq:G:complet} and \ref{eq:sigma:complet}:
\begin{align}
    g(t_a)=-\frac{uV^*}{V}\left(e^s-1\right)\\
    G=\frac{V^*}{V}\int_0^\infty ds \exp{\left(-\frac{uV^*}{V}(e^s-1)\right)}\\
     \sigma_t=\frac{kT}{\lambda}\frac{V^*}{V}\frac{N}{G}\int_0^\infty s ds~\exp{\left(-\frac{uV^*}{V}(e^s-1)\right)}
\end{align}
Here, we introduce the ratio $\hat{V}=V/V^*$ so that $s=t_a/\tau \hat{V}$
. It follows:
\begin{align}
 g(t_a)&=\exp\left[ -\frac u{\hat{V}}\left( \exp\left(\frac{\hat{V}t_a}\tau-1  \right)\right)\right] \\
 G&=\frac{\tau}{\hat{V}}\exp\left( \frac{u}{\hat{V} }\right)E_1\left( \frac{u}{\hat{V}} \right)
\end{align} 
where we use the exponential integral
\begin{equation}
 E_1\left( x \right)=\int_x^\infty \frac{e^{-y}}{y}dy
\end{equation} 
The calculation arrives at:
\begin{equation}
 \sigma_t=\frac{kTN_0}{\lambda}\frac{\frac{1}{\hat{V}}\exp\left( \frac{u}{\hat{V}} \right)}{1+\frac{1}{\hat{V}_0}\exp\left( \frac{u}{\hat{V}} \right)E_1\left( \frac{u}{\hat{V}} \right)}
 \left[ G_1\left( \frac{u}{\hat{V}} \right)-\ln\left(  \frac{u}{\hat{V}}  \right)E_1\left( \frac{u}{\hat{V}} \right)
 \right] 
\end{equation} 
with, using Maple,
\footnote[2]{
This result can also be derived using DLMF \cite{DLMF}
$${{}_{p+1}F_{q+1}}\left({a_{0},\dots,a_{p}\atop b_{0},\dots,b_{q}};z\right)=$$\\
$$\frac{\Gamma\left(b_{0}\right)}{\Gamma\left(a_{0}\right)\Gamma\left(b_{0}-a_{0%
}\right)}\int_{0}^{1}t^{a_{0}-1}(1-t)^{b_{0}-a_{0}-1}{{}_{p}F_{q}}\left({a_{1}
,\dots,a_{p}\atop b_{1},\dots,b_{q}};zt\right)\mathrm{d}t$$
to express
$${{}_3F_3}\left({1,1,1\atop 2,2,2};z\right)$$
from Wolfram \cite{Wolfram}\\
${ }_{2} F_{2}(1,1 ; 2,2 ;
z)=\frac{1}{z}\left(\operatorname{Ei}(z)+\frac{1}{2}\left(\log
\left(\frac{1}{z}\right)-\log (z)\right)-\gamma\right)$
} 

\begin{align}
 G_1\left( x \right)= & \int_x^\infty \frac{e^{-y}}{y}\ln y dy\\
= & \ln\left( x \right)E_1\left( x \right)+x\Bigg[ \frac{1}{2x}\left( 2\gamma\ln\left( x \right)+\ln^2\left( x \right)+\gamma^2+\frac{\pi^2}{6} \right) \\
&  -{}_3F_3\left( 1,1,1;2,2,2;-x \right)\Bigg]
\end{align} 
Finally,
\begin{multline}
    \sigma_t=\frac{kTN_0}{2\lambda}\frac{\frac{1}{\hat{V}}\exp\left( \frac{u}{\hat{V}} \right)}{1+\frac{1}{\hat{V}_0}\exp\left( \frac{u}{\hat{V}} \right)E_1\left( \frac{u}{\hat{V}} \right)} \\
\left[ \left(\gamma + \ln\frac{u}{\hat{V}}\right)^2+\frac{\pi^2}{6}
-\frac{2u}{\hat{V}}\,_3F_3\left( 1,1,1;2,2,2;-\frac{u}{\hat{V}} \right)\right]
\label{eq:sigcompl}
\end{multline} 
The variations of $\sigma_t$ with $\hat{V}$ are plotted in Figure~\ref{Fig:singh} as a lin-log plot and show the well-known bell-shaped curve with an increasing logarithmic branch between $uV^*$ and $V^*$.  {Similar equations were derived in the context of friction of adhesive patches of SAMs.\cite{drummond_friction_2003}}\\


\bibliography{biblio} 

\providecommand*{\mcitethebibliography}{\thebibliography}
\csname @ifundefined\endcsname{endmcitethebibliography}
{\let\endmcitethebibliography\endthebibliography}{}
\begin{mcitethebibliography}{50}
\providecommand*{\natexlab}[1]{#1}
\providecommand*{\mciteSetBstSublistMode}[1]{}
\providecommand*{\mciteSetBstMaxWidthForm}[2]{}
\providecommand*{\mciteBstWouldAddEndPuncttrue}
  {\def\EndOfBibitem{\unskip.}}
\providecommand*{\mciteBstWouldAddEndPunctfalse}
  {\let\EndOfBibitem\relax}
\providecommand*{\mciteSetBstMidEndSepPunct}[3]{}
\providecommand*{\mciteSetBstSublistLabelBeginEnd}[3]{}
\providecommand*{\EndOfBibitem}{}
\mciteSetBstSublistMode{f}
\mciteSetBstMaxWidthForm{subitem}
{(\emph{\alph{mcitesubitemcount}})}
\mciteSetBstSublistLabelBeginEnd{\mcitemaxwidthsubitemform\space}
{\relax}{\relax}

\bibitem[Porte \emph{et~al.}(2020)Porte, Cann, and
  Masen]{porte_lubrication_2020}
E.~Porte, P.~Cann and M.~Masen, \emph{Soft Matter}, 2020, \textbf{16},
  10290--10300\relax
\mciteBstWouldAddEndPuncttrue
\mciteSetBstMidEndSepPunct{\mcitedefaultmidpunct}
{\mcitedefaultendpunct}{\mcitedefaultseppunct}\relax
\EndOfBibitem
\bibitem[Blum and Ovaert(2013)]{blum_low_2013}
M.~M. Blum and T.~C. Ovaert, \emph{Materials Science and Engineering: C}, 2013,
  \textbf{33}, 4377--4383\relax
\mciteBstWouldAddEndPuncttrue
\mciteSetBstMidEndSepPunct{\mcitedefaultmidpunct}
{\mcitedefaultendpunct}{\mcitedefaultseppunct}\relax
\EndOfBibitem
\bibitem[Han and Eriten(2018)]{han_effect_2018}
G.~Han and M.~Eriten, \emph{Royal Society Open Science}, 2018, \textbf{5},
  172051\relax
\mciteBstWouldAddEndPuncttrue
\mciteSetBstMidEndSepPunct{\mcitedefaultmidpunct}
{\mcitedefaultendpunct}{\mcitedefaultseppunct}\relax
\EndOfBibitem
\bibitem[Dunn \emph{et~al.}(2013)Dunn, Urueña, Huo, Perry, Angelini, and
  Sawyer]{dunn_lubricity_2013}
A.~C. Dunn, J.~M. Urueña, Y.~Huo, S.~S. Perry, T.~E. Angelini and W.~G.
  Sawyer, \emph{Tribology Letters}, 2013, \textbf{49}, 371--378\relax
\mciteBstWouldAddEndPuncttrue
\mciteSetBstMidEndSepPunct{\mcitedefaultmidpunct}
{\mcitedefaultendpunct}{\mcitedefaultseppunct}\relax
\EndOfBibitem
\bibitem[Roba \emph{et~al.}(2011)Roba, Duncan, Hill, Spencer, and
  Tosatti]{roba_friction_2011}
M.~Roba, E.~G. Duncan, G.~A. Hill, N.~D. Spencer and S.~G.~P. Tosatti,
  \emph{Tribology Letters}, 2011, \textbf{44}, 387--397\relax
\mciteBstWouldAddEndPuncttrue
\mciteSetBstMidEndSepPunct{\mcitedefaultmidpunct}
{\mcitedefaultendpunct}{\mcitedefaultseppunct}\relax
\EndOfBibitem
\bibitem[Delavoipière \emph{et~al.}(2018)Delavoipière, Heurtefeu, Teisseire,
  Chateauminois, Tran, Fermigier, and Verneuil]{delavoipiere_swelling_2018}
J.~Delavoipière, B.~Heurtefeu, J.~Teisseire, A.~Chateauminois, Y.~Tran,
  M.~Fermigier and E.~Verneuil, \emph{Langmuir}, 2018, \textbf{34},
  15238--15244\relax
\mciteBstWouldAddEndPuncttrue
\mciteSetBstMidEndSepPunct{\mcitedefaultmidpunct}
{\mcitedefaultendpunct}{\mcitedefaultseppunct}\relax
\EndOfBibitem
\bibitem[Lee \emph{et~al.}(2013)Lee, Alcaraz, Rubner, and
  Cohen]{lee_zwitter-wettability_2013}
H.~Lee, M.~L. Alcaraz, M.~F. Rubner and R.~E. Cohen, \emph{ACS Nano}, 2013,
  \textbf{7}, 2172--2185\relax
\mciteBstWouldAddEndPuncttrue
\mciteSetBstMidEndSepPunct{\mcitedefaultmidpunct}
{\mcitedefaultendpunct}{\mcitedefaultseppunct}\relax
\EndOfBibitem
\bibitem[Tominaga \emph{et~al.}(2008)Tominaga, Takedomi, Biederman, Furukawa,
  Osada, and Gong]{tominaga_effect_2008}
T.~Tominaga, N.~Takedomi, H.~Biederman, H.~Furukawa, Y.~Osada and J.~P. Gong,
  \emph{Soft Matter}, 2008, \textbf{4}, 1033\relax
\mciteBstWouldAddEndPuncttrue
\mciteSetBstMidEndSepPunct{\mcitedefaultmidpunct}
{\mcitedefaultendpunct}{\mcitedefaultseppunct}\relax
\EndOfBibitem
\bibitem[Kagata \emph{et~al.}(2001)Kagata, Gong, and
  Osada]{kagata_surface_2001}
G.~Kagata, J.~P. Gong and Y.~Osada, \emph{Wear}, 2001, \textbf{251},
  1188--1192\relax
\mciteBstWouldAddEndPuncttrue
\mciteSetBstMidEndSepPunct{\mcitedefaultmidpunct}
{\mcitedefaultendpunct}{\mcitedefaultseppunct}\relax
\EndOfBibitem
\bibitem[Schallamach(1963)]{schallamach_theory_1963}
A.~Schallamach, \emph{Wear}, 1963, \textbf{6}, 375--382\relax
\mciteBstWouldAddEndPuncttrue
\mciteSetBstMidEndSepPunct{\mcitedefaultmidpunct}
{\mcitedefaultendpunct}{\mcitedefaultseppunct}\relax
\EndOfBibitem
\bibitem[Skotheim and Mahadevan(2005)]{skotheim_soft_2005}
J.~M. Skotheim and L.~Mahadevan, \emph{Physics of Fluids}, 2005, \textbf{17},
  092101\relax
\mciteBstWouldAddEndPuncttrue
\mciteSetBstMidEndSepPunct{\mcitedefaultmidpunct}
{\mcitedefaultendpunct}{\mcitedefaultseppunct}\relax
\EndOfBibitem
\bibitem[Salez and Mahadevan(2015)]{salez_elastohydrodynamics_2015}
T.~Salez and L.~Mahadevan, \emph{Journal of Fluid Mechanics}, 2015,
  \textbf{779}, 181--196\relax
\mciteBstWouldAddEndPuncttrue
\mciteSetBstMidEndSepPunct{\mcitedefaultmidpunct}
{\mcitedefaultendpunct}{\mcitedefaultseppunct}\relax
\EndOfBibitem
\bibitem[Gong and Osada(1998)]{gong_gel_1998}
J.~Gong and Y.~Osada, \emph{The Journal of Chemical Physics}, 1998,
  \textbf{109}, 8062--8068\relax
\mciteBstWouldAddEndPuncttrue
\mciteSetBstMidEndSepPunct{\mcitedefaultmidpunct}
{\mcitedefaultendpunct}{\mcitedefaultseppunct}\relax
\EndOfBibitem
\bibitem[Xiang \emph{et~al.}(2020)Xiang, Zhang, Gong, and
  Zeng]{xiang_surface_2020}
L.~Xiang, J.~Zhang, L.~Gong and H.~Zeng, \emph{Soft Matter}, 2020, \textbf{16},
  6697--6719\relax
\mciteBstWouldAddEndPuncttrue
\mciteSetBstMidEndSepPunct{\mcitedefaultmidpunct}
{\mcitedefaultendpunct}{\mcitedefaultseppunct}\relax
\EndOfBibitem
\bibitem[Yamamoto \emph{et~al.}(2014)Yamamoto, Kurokawa, Ahmed, Kamita,
  Yashima, Furukawa, Ota, Furukawa, and Gong]{yamamoto_situ_2014}
T.~Yamamoto, T.~Kurokawa, J.~Ahmed, G.~Kamita, S.~Yashima, Y.~Furukawa, Y.~Ota,
  H.~Furukawa and J.~P. Gong, \emph{Soft Matter}, 2014, \textbf{10},
  5589--5596\relax
\mciteBstWouldAddEndPuncttrue
\mciteSetBstMidEndSepPunct{\mcitedefaultmidpunct}
{\mcitedefaultendpunct}{\mcitedefaultseppunct}\relax
\EndOfBibitem
\bibitem[Simi{\v c} \emph{et~al.}(2020)Simi{\v c}, Yetkin, Zhang, and
  Spencer]{simic_importance_2020}
R.~Simi{\v c}, M.~Yetkin, K.~Zhang and N.~D. Spencer, \emph{Tribology Letters},
  2020, \textbf{68}, 64\relax
\mciteBstWouldAddEndPuncttrue
\mciteSetBstMidEndSepPunct{\mcitedefaultmidpunct}
{\mcitedefaultendpunct}{\mcitedefaultseppunct}\relax
\EndOfBibitem
\bibitem[Baumberger \emph{et~al.}(2003)Baumberger, Caroli, and
  Ronsin]{baumberger_self-healing_2003}
T.~Baumberger, C.~Caroli and O.~Ronsin, \emph{The European Physical Journal E},
  2003, \textbf{11}, 85--93\relax
\mciteBstWouldAddEndPuncttrue
\mciteSetBstMidEndSepPunct{\mcitedefaultmidpunct}
{\mcitedefaultendpunct}{\mcitedefaultseppunct}\relax
\EndOfBibitem
\bibitem[Kagata \emph{et~al.}(2002)Kagata, Gong, and
  Osada]{kagata_friction_2002}
G.~Kagata, J.~P. Gong and Y.~Osada, \emph{The Journal of Physical Chemistry B},
  2002, \textbf{106}, 4596--4601\relax
\mciteBstWouldAddEndPuncttrue
\mciteSetBstMidEndSepPunct{\mcitedefaultmidpunct}
{\mcitedefaultendpunct}{\mcitedefaultseppunct}\relax
\EndOfBibitem
\bibitem[Kurokawa \emph{et~al.}(2005)Kurokawa, Tominaga, Katsuyama, Kuwabara,
  Furukawa, Osada, and Gong]{kurokawa_elastichydrodynamic_2005}
T.~Kurokawa, T.~Tominaga, Y.~Katsuyama, R.~Kuwabara, H.~Furukawa, Y.~Osada and
  J.~P. Gong, \emph{Langmuir}, 2005, \textbf{21}, 8643--8648\relax
\mciteBstWouldAddEndPuncttrue
\mciteSetBstMidEndSepPunct{\mcitedefaultmidpunct}
{\mcitedefaultendpunct}{\mcitedefaultseppunct}\relax
\EndOfBibitem
\bibitem[Cuccia \emph{et~al.}(2020)Cuccia, Pothineni, Wu, Méndez~Harper, and
  Burton]{cuccia_pore-size_2020}
N.~L. Cuccia, S.~Pothineni, B.~Wu, J.~Méndez~Harper and J.~C. Burton,
  \emph{Proceedings of the National Academy of Sciences}, 2020, \textbf{117},
  11247--11256\relax
\mciteBstWouldAddEndPuncttrue
\mciteSetBstMidEndSepPunct{\mcitedefaultmidpunct}
{\mcitedefaultendpunct}{\mcitedefaultseppunct}\relax
\EndOfBibitem
\bibitem[Delavoipière \emph{et~al.}(2018)Delavoipière, Tran, Verneuil,
  Heurtefeu, Hui, and Chateauminois]{delavoipiere_friction_2018}
J.~Delavoipière, Y.~Tran, E.~Verneuil, B.~Heurtefeu, C.~Y. Hui and
  A.~Chateauminois, \emph{Langmuir}, 2018, \textbf{34}, 9617--9626\relax
\mciteBstWouldAddEndPuncttrue
\mciteSetBstMidEndSepPunct{\mcitedefaultmidpunct}
{\mcitedefaultendpunct}{\mcitedefaultseppunct}\relax
\EndOfBibitem
\bibitem[Ciapa \emph{et~al.}(2020)Ciapa, Delavoipi{\`e}re, Tran, Verneuil, and
  Chateauminois]{ciapa_transient_2020}
L.~Ciapa, J.~Delavoipi{\`e}re, Y.~Tran, E.~Verneuil and A.~Chateauminois,
  \emph{Soft Matter}, 2020, \textbf{16}, 6539--6548\relax
\mciteBstWouldAddEndPuncttrue
\mciteSetBstMidEndSepPunct{\mcitedefaultmidpunct}
{\mcitedefaultendpunct}{\mcitedefaultseppunct}\relax
\EndOfBibitem
\bibitem[Shoaib and Espinosa-Marzal(2018)]{shoaib_insight_2018}
T.~Shoaib and R.~M. Espinosa-Marzal, \emph{Tribology Letters}, 2018,
  \textbf{66}, 96\relax
\mciteBstWouldAddEndPuncttrue
\mciteSetBstMidEndSepPunct{\mcitedefaultmidpunct}
{\mcitedefaultendpunct}{\mcitedefaultseppunct}\relax
\EndOfBibitem
\bibitem[Johnson \emph{et~al.}(1971)Johnson, Kendall, and
  Roberts]{johnson_surface_1971}
K.~Johnson, K.~Kendall and A.~Roberts, \emph{Proceedings of the Royal Society
  of London Series A-Mathematical and Physical sciences}, 1971, \textbf{324},
  301\relax
\mciteBstWouldAddEndPuncttrue
\mciteSetBstMidEndSepPunct{\mcitedefaultmidpunct}
{\mcitedefaultendpunct}{\mcitedefaultseppunct}\relax
\EndOfBibitem
\bibitem[H\'enot \emph{et~al.}(2018)H\'enot, Grzelka, Zhang, Mariot, Antoniuk,
  Drockenmuller, L\'eger, and Restagno]{henot_temperature_2018}
M.~H\'enot, M.~Grzelka, J.~Zhang, S.~Mariot, I.~Antoniuk, E.~Drockenmuller,
  L.~L\'eger and F.~Restagno, \emph{Phys. Rev. Lett.}, 2018, \textbf{121},
  177802\relax
\mciteBstWouldAddEndPuncttrue
\mciteSetBstMidEndSepPunct{\mcitedefaultmidpunct}
{\mcitedefaultendpunct}{\mcitedefaultseppunct}\relax
\EndOfBibitem
\bibitem[Henot \emph{et~al.}(2018)Henot, Drockenmuller, Leger, and
  Restagno]{henot_friction_2018}
M.~Henot, E.~Drockenmuller, L.~Leger and F.~Restagno, \emph{Acs Macro Letters},
  2018, \textbf{7}, 112--115\relax
\mciteBstWouldAddEndPuncttrue
\mciteSetBstMidEndSepPunct{\mcitedefaultmidpunct}
{\mcitedefaultendpunct}{\mcitedefaultseppunct}\relax
\EndOfBibitem
\bibitem[Bureau(2007)]{bureau_friction_2007}
L.~Bureau, \emph{Macromolecules}, 2007, \textbf{40}, 9197--9200\relax
\mciteBstWouldAddEndPuncttrue
\mciteSetBstMidEndSepPunct{\mcitedefaultmidpunct}
{\mcitedefaultendpunct}{\mcitedefaultseppunct}\relax
\EndOfBibitem
\bibitem[Bozna \emph{et~al.}(2015)Bozna, Blass, Albrecht, Hausen, Wenz, and
  Bennewitz]{bozna_friction_2015}
B.~L. Bozna, J.~Blass, M.~Albrecht, F.~Hausen, G.~Wenz and R.~Bennewitz,
  \emph{Langmuir}, 2015, \textbf{31}, 10708--10716\relax
\mciteBstWouldAddEndPuncttrue
\mciteSetBstMidEndSepPunct{\mcitedefaultmidpunct}
{\mcitedefaultendpunct}{\mcitedefaultseppunct}\relax
\EndOfBibitem
\bibitem[Blass \emph{et~al.}(2015)Blass, Bozna, Albrecht, Krings, Ravoo, Wenz,
  and Bennewitz]{blass_switching_2015}
J.~Blass, B.~L. Bozna, M.~Albrecht, J.~A. Krings, B.~J. Ravoo, G.~Wenz and
  R.~Bennewitz, \emph{Chemical Communications}, 2015, \textbf{51},
  1830--1833\relax
\mciteBstWouldAddEndPuncttrue
\mciteSetBstMidEndSepPunct{\mcitedefaultmidpunct}
{\mcitedefaultendpunct}{\mcitedefaultseppunct}\relax
\EndOfBibitem
\bibitem[Friddle \emph{et~al.}(2008)Friddle, Podsiadlo, Artyukhin, and
  Noy]{friddle_near-equilibrium_2008}
R.~W. Friddle, P.~Podsiadlo, A.~B. Artyukhin and A.~Noy, \emph{The Journal of
  Physical Chemistry C}, 2008, \textbf{112}, 4986--4990\relax
\mciteBstWouldAddEndPuncttrue
\mciteSetBstMidEndSepPunct{\mcitedefaultmidpunct}
{\mcitedefaultendpunct}{\mcitedefaultseppunct}\relax
\EndOfBibitem
\bibitem[Kühner \emph{et~al.}(2006)Kühner, Erdmann, Sonnenberg, Serr,
  Morfill, and Gaub]{kuhner_friction_2006}
F.~Kühner, M.~Erdmann, L.~Sonnenberg, A.~Serr, J.~Morfill and H.~E. Gaub,
  \emph{Langmuir}, 2006, \textbf{22}, 11180--11186\relax
\mciteBstWouldAddEndPuncttrue
\mciteSetBstMidEndSepPunct{\mcitedefaultmidpunct}
{\mcitedefaultendpunct}{\mcitedefaultseppunct}\relax
\EndOfBibitem
\bibitem[Drummond \emph{et~al.}(2003)Drummond, Israelachvili, and
  Richetti]{drummond_friction_2003}
C.~Drummond, J.~Israelachvili and P.~Richetti, \emph{Physical Review E}, 2003,
  \textbf{67}, 066110\relax
\mciteBstWouldAddEndPuncttrue
\mciteSetBstMidEndSepPunct{\mcitedefaultmidpunct}
{\mcitedefaultendpunct}{\mcitedefaultseppunct}\relax
\EndOfBibitem
\bibitem[Delavoipi{\`e}re \emph{et~al.}(2016)Delavoipi{\`e}re, Tran, Verneuil,
  and Chateauminois]{delavoipiere_poroelastic_2016}
J.~Delavoipi{\`e}re, Y.~Tran, E.~Verneuil and A.~Chateauminois, \emph{Soft
  Matter}, 2016, \textbf{12}, 8049--8058\relax
\mciteBstWouldAddEndPuncttrue
\mciteSetBstMidEndSepPunct{\mcitedefaultmidpunct}
{\mcitedefaultendpunct}{\mcitedefaultseppunct}\relax
\EndOfBibitem
\bibitem[Augustine \emph{et~al.}(2023)Augustine, Veillerot, Gauthier, Zhu, Hui,
  Tran, Verneuil, and Chateauminois]{augustine_swelling_2023}
A.~Augustine, M.~Veillerot, N.~Gauthier, B.~Zhu, C.-Y. Hui, Y.~Tran,
  E.~Verneuil and A.~Chateauminois, \emph{Soft Matter}, 2023, \textbf{19},
  5169--5178\relax
\mciteBstWouldAddEndPuncttrue
\mciteSetBstMidEndSepPunct{\mcitedefaultmidpunct}
{\mcitedefaultendpunct}{\mcitedefaultseppunct}\relax
\EndOfBibitem
\bibitem[Wang and Lieberman(2003)]{wang_growth_2003}
Y.~Wang and M.~Lieberman, \emph{Langmuir}, 2003, \textbf{19}, 1159--1167\relax
\mciteBstWouldAddEndPuncttrue
\mciteSetBstMidEndSepPunct{\mcitedefaultmidpunct}
{\mcitedefaultendpunct}{\mcitedefaultseppunct}\relax
\EndOfBibitem
\bibitem[Brzoska \emph{et~al.}(1994)Brzoska, Azouz, and
  Rondelez]{brzoska_silanization_1994}
J.~B. Brzoska, I.~B. Azouz and F.~Rondelez, \emph{Langmuir}, 1994, \textbf{10},
  4367--4373\relax
\mciteBstWouldAddEndPuncttrue
\mciteSetBstMidEndSepPunct{\mcitedefaultmidpunct}
{\mcitedefaultendpunct}{\mcitedefaultseppunct}\relax
\EndOfBibitem
\bibitem[Vandenberg \emph{et~al.}(1991)Vandenberg, Bertilsson, Liedberg, Uvdal,
  Erlandsson, Elwing, and Lundstr{\"o}m]{vandenberg_structure_1991}
E.~T. Vandenberg, L.~Bertilsson, B.~Liedberg, K.~Uvdal, R.~Erlandsson,
  H.~Elwing and I.~Lundstr{\"o}m, \emph{Journal of Colloid and Interface
  Science}, 1991, \textbf{147}, 103--118\relax
\mciteBstWouldAddEndPuncttrue
\mciteSetBstMidEndSepPunct{\mcitedefaultmidpunct}
{\mcitedefaultendpunct}{\mcitedefaultseppunct}\relax
\EndOfBibitem
\bibitem[Theodoly \emph{et~al.}(2010)Theodoly, Huang, and
  Valignat]{theodoly2010}
O.~Theodoly, Z.-H. Huang and M.-P. Valignat, \emph{Langmuir}, 2010,
  \textbf{26}, 1940--1948\relax
\mciteBstWouldAddEndPuncttrue
\mciteSetBstMidEndSepPunct{\mcitedefaultmidpunct}
{\mcitedefaultendpunct}{\mcitedefaultseppunct}\relax
\EndOfBibitem
\bibitem[Chernyak and Leonov(1986)]{chernyak_theory_1986}
Y.~Chernyak and A.~Leonov, \emph{Wear}, 1986, \textbf{108}, 105--138\relax
\mciteBstWouldAddEndPuncttrue
\mciteSetBstMidEndSepPunct{\mcitedefaultmidpunct}
{\mcitedefaultendpunct}{\mcitedefaultseppunct}\relax
\EndOfBibitem
\bibitem[Leonov(1990)]{leonov_dependence_1990}
A.~Leonov, \emph{Wear}, 1990, \textbf{141}, 137--145\relax
\mciteBstWouldAddEndPuncttrue
\mciteSetBstMidEndSepPunct{\mcitedefaultmidpunct}
{\mcitedefaultendpunct}{\mcitedefaultseppunct}\relax
\EndOfBibitem
\bibitem[Singh and Juvekar(2011)]{singh_steady_2011}
A.~K. Singh and V.~A. Juvekar, \emph{Soft Matter}, 2011, \textbf{7},
  10601\relax
\mciteBstWouldAddEndPuncttrue
\mciteSetBstMidEndSepPunct{\mcitedefaultmidpunct}
{\mcitedefaultendpunct}{\mcitedefaultseppunct}\relax
\EndOfBibitem
\bibitem[Friddle \emph{et~al.}(2012)Friddle, Noy, and
  De~Yoreo]{friddle_interpreting_2012}
R.~W. Friddle, A.~Noy and J.~J. De~Yoreo, \emph{Proceedings of the National
  Academy of Sciences}, 2012, \textbf{109}, 13573--13578\relax
\mciteBstWouldAddEndPuncttrue
\mciteSetBstMidEndSepPunct{\mcitedefaultmidpunct}
{\mcitedefaultendpunct}{\mcitedefaultseppunct}\relax
\EndOfBibitem
\bibitem[Eyring(1936)]{eyring_viscosity_1936}
H.~Eyring, \emph{The Journal of Chemical Physics}, 1936, \textbf{4},
  283--291\relax
\mciteBstWouldAddEndPuncttrue
\mciteSetBstMidEndSepPunct{\mcitedefaultmidpunct}
{\mcitedefaultendpunct}{\mcitedefaultseppunct}\relax
\EndOfBibitem
\bibitem[Rubinstein and Colby(2003)]{rubinstein_polymer_2003}
M.~Rubinstein and R.~Colby, \emph{Polymer {Physics}}, Oxford University Press,
  2003\relax
\mciteBstWouldAddEndPuncttrue
\mciteSetBstMidEndSepPunct{\mcitedefaultmidpunct}
{\mcitedefaultendpunct}{\mcitedefaultseppunct}\relax
\EndOfBibitem
\bibitem[Vorvolakos and Chaudhury(2003)]{vorvolakos_effects_2003}
K.~Vorvolakos and M.~Chaudhury, \emph{LANGMUIR}, 2003, \textbf{19},
  6778--6787\relax
\mciteBstWouldAddEndPuncttrue
\mciteSetBstMidEndSepPunct{\mcitedefaultmidpunct}
{\mcitedefaultendpunct}{\mcitedefaultseppunct}\relax
\EndOfBibitem
\bibitem[Kékicheff(2019)]{kekicheff_long-range_2019}
P.~Kékicheff, \emph{Advances in Colloid and Interface Science}, 2019,
  \textbf{270}, 191--215\relax
\mciteBstWouldAddEndPuncttrue
\mciteSetBstMidEndSepPunct{\mcitedefaultmidpunct}
{\mcitedefaultendpunct}{\mcitedefaultseppunct}\relax
\EndOfBibitem
\bibitem[Shen \emph{et~al.}(2019)Shen, Jiang, Cao, and Zhang]{shen_effect_2019}
L.~Shen, H.~Jiang, J.~Cao and H.~Zhang, \emph{Construction and Building
  Materials}, 2019, \textbf{214}, 101--110\relax
\mciteBstWouldAddEndPuncttrue
\mciteSetBstMidEndSepPunct{\mcitedefaultmidpunct}
{\mcitedefaultendpunct}{\mcitedefaultseppunct}\relax
\EndOfBibitem
\bibitem[Singh \emph{et~al.}(2021)Singh, Juvekar, and
  Bellare]{singh_model_2021}
A.~K. Singh, V.~A. Juvekar and J.~R. Bellare, \emph{Journal of the Mechanics
  and Physics of Solids}, 2021, \textbf{146}, 104191\relax
\mciteBstWouldAddEndPuncttrue
\mciteSetBstMidEndSepPunct{\mcitedefaultmidpunct}
{\mcitedefaultendpunct}{\mcitedefaultseppunct}\relax
\EndOfBibitem
\bibitem[DLM()]{DLMF}
\emph{{Digital Library of Mathematical Functions}},
  \url{https://dlmf.nist.gov/16.5.E2}, Wolfram\relax
\mciteBstWouldAddEndPuncttrue
\mciteSetBstMidEndSepPunct{\mcitedefaultmidpunct}
{\mcitedefaultendpunct}{\mcitedefaultseppunct}\relax
\EndOfBibitem
\bibitem[Wol()]{Wolfram}
\emph{{Generalized Hypergeometric Function}},
  \url{http://functions.wolfram.com/07.31.03.0120.01}, Wolfram\relax
\mciteBstWouldAddEndPuncttrue
\mciteSetBstMidEndSepPunct{\mcitedefaultmidpunct}
{\mcitedefaultendpunct}{\mcitedefaultseppunct}\relax
\EndOfBibitem
\end{mcitethebibliography}
\bibliographystyle{rsc} 

\end{document}